**Unexpected antagonism between ferroelectricity and Rashba effects in epitaxially strained SrTiO$_3$**

Julien Varignon[1]

[1] CRISMAT, ENSICAEN, Normandie Université, UNICAEN, CNRS, 14000 Caen, France

**Abstract**

The Rashba parameter $\alpha_R$ is usually assumed to scale linearly with the amplitude of polar displacements by construction of the spin-orbit interaction. On the basis of *first-principles* simulations, ferroelectric phases of SrTiO$_3$ reached under epitaxial compressive strain are characterized by (i) large Rashba effects at the bottom of the conduction band near the paraelectric-ferroelectric boundary and (ii) an unexpected suppression of the phenomena when the amplitude of polar displacements increases. This peculiar behavior is ascribed to the inverse dependance of the Rashba parameter with the crystal field $\Delta_{CF}$ induced by the polar displacements that alleviates the degeneracy of Ti $t_{2g}$ states and annihilates the Rashba effects. Although $\alpha_R$ has intrinsically a linear dependance on polar displacements, these latter becomes antagonist to Rashba phenomena at large polar mode amplitude. Thus, the Rashba coefficient may be bound to a upper value.

Spintronic exploits the spin degree of freedom in addition to the charge of carriers, yielding numerous applications in data storage [1]. Traditionally, spintronic relies on ferromagnetic metals such as Ni or Co as generators and detectors of spin currents. Nevertheless, ferromagnetic metals have several drawbacks: (i) they can generate undesired local magnetic fields hindering the densification of devices and (ii) a large voltage is required for reversing their magnetization, hence incompatible with our quest for low energy consumption devices. An alternative pathway toward lower power spintronics exists and exploits the spin-orbit interaction (SOI) of non-magnetic materials through the Spin Hall Effect (SHE) [2,3] and inverse Spin Hall Effect (ISHE) [4,5], alleviating the use of ferromagnets for generating spin currents [6].

More recently, a promising pathway toward efficient spin-charge current interconversion has been identified and uses a peculiar interplay between polar displacements and SOI: when inversion symmetry is broken, such as at interfaces, surfaces or in ferroelectric compounds, the polar displacements can yield a Rashba interaction lifting the degeneracy of bands according to their spin [7]. For a polar axis along z, the Rashba SOI takes the following expression

$$H_R = \alpha_R (S_y k_x - S_x k_y)$$ **(eq.1)**

where $\alpha_R$ is the Rashba parameter ($\alpha_R \propto Q_P$ where $Q_P$ is the polar displacement amplitude in the material), $k_i$ is the momentum of electrons and $S_i$ is the spin direction (i=x,y). A spin locking is thus achieved with spins of carriers being orthogonal to the momentum k. Spin-to-charge current interconversions are then enabled through the Edelstein and inverse Edelstein effects [8,9], whose efficiency directly scales with the amplitude of the Rashba parameter [9].

At first glance, reaching a large Rashba phenomenon requires two key ingredients: (i) the material has to exhibit a sizable intrinsic SOI – *i.e.* large $\lambda_{SO}$ – and (ii) the material must display large polar displacements amplitude. The latter criterion induces that $\alpha_R$ should scale linearly with the polar distortion amplitude while the former is usually reached for heavy elements and/or for compounds with highly degenerate states [10–12]. At interfaces or surfaces, polar displacements are pinned by the structure but ferroelectric materials offer a non-volatile control of $\alpha_R$ through the reversal of polar displacements with an external electric field. However, structural displacements such as breathing distortions in $SrBiO_3$ associated with charge orderings [13] or octahedral rotations appearing in ferroelectric oxides [14] can tune the Rashba parameter. One may thus question if "*the Rashba parameter always scales linearly with the polar displacements amplitude?*" and "*if the related structural displacements can tune $\alpha_R$ in a different manner?*". It is important to clarify these questions as it could possibly unveil strategies to design compounds with larger Rashba parameters and possibly more efficient current interconversions.

In that context, strontium titanate perovskite oxide SrTiO$_3$ (STO) is a prototypical material for exploring the interplay between polar displacements and Rashba effects since (i) it enables efficient spin-to-charge current interconversion in the bi-dimensional electron gas (2-DEG) formed at its interface with LaAlO$_3$ [15] – albeit the Rashba parameter is small, the efficiency of the interconversion also directly rely on the carrier lifetime that is large in STO – and (ii) oxide perovskites can be turned into ferroelectrics using various external stimuli [16]. In the bulk ground state, STO adopts a *I$_4$/mcm* tetragonal symmetry characterized by a $a^0a^0c^-$ O$_6$ groups rotations following Glazer's notations [17] – also called antiferrodistortive (AFD) motion – with respect to the perfectly undistorted *Pm-3m* cubic cell existing at higher temperatures (**Figure 1.a**). STO is a paraelectric at all temperatures but it becomes ferroelectric with a moderate epitaxial strain [18]. However, the strain phase diagram of STO is up to date rather elusive, hindering the identification of potentially interesting ferroelectric phases for spin-orbitronic applications.

In this letter, I explore using *first-principles* simulations the phase diagram of SrTiO$_3$ experiencing compressive epitaxial strain and I reveal the existence of polar ground states for strain values $\eta$ larger than 1.35% with respect to the bulk lattice parameter. The spontaneous polarization increases with increasing $\eta$ until reaching a super-tetragonal phase with a *c/a* ratio of 1.29 and a polarization approaching 100 µC.cm$^{-2}$. Within the ferroelectric phases, a Rashba spin splitting of the Ti $t_{2g}$ bands located at the bottom of the conduction band is identified but the amplitude of the Rashba parameter is surprisingly found to diminish with the polar displacements amplitude. This is due to the fact that polar displacements alleviate the Ti $t_{2g}$ orbitals degeneracies thereby weakening the Rashba-SOI interaction. Ultimately, the phases with the largest polarizations do not show any Rashba spin-splitting at the bottom of the conduction band. Thus, the interplay between polar displacements and the Rashba parameter is subtle in SrTiO$_3$ and preserving degenerate states rather than reaching high electrical dipoles might be a better route toward maximal Rashba parameters and more efficient spin-charge current interconversion.

***Method:*** *First-principles* Density Functional Theory (DFT) calculations are performed with the Vienna Ab initio Simulation Package (VASP [19,20]) using the meta-Generalized Gradient Approximation (meta-GGA) Strongly Constrained and Appropriately Normalized (SCAN) exchange-correlation functional [21]. The SCAN functional was previously shown to be well suited for the study of perovskite oxides [22] and of ferroic properties of ATiO$_3$ compounds (A=Ba, Sr, Pb) [23], including the paraelectric nature of bulk STO (see **Supplementary Material 1** [16,17,31–34,23–30]). Epitaxial strain is modelled by imposing two lattice parameters to that of the substrate. Then, the third lattice parameter is allowed to relax in amplitude and direction and atomic positions are totally optimized until forces acting on each atom are lower than 0.001 eV/Å. In order to avoid getting trapped in false local minimums during geometry optimizations, several structures exhibiting (i) octahedral tilt patterns such as $a^0a^0c^0$, $a^0a^0c^-$, $a^-a^0c^0$, $a^0a^0c^+$, $a^+a^0c^0$, $a^-a^-c^+$, $a^+a^-c^-$ or $a^-a^-c^-$ in Glazer's notation [35] as well as (ii) two polar axis either along or orthogonal to the substrate or no polar axis are considered as initial points for the structural relaxations. All relaxations are performed without the SOI, this interaction is only considered afterwards for band structure calculations. The cut-off is set to 650 eV and is accompanied by a 8x8x6 and 6x6x6 kmesh for ($2\sqrt{2}$, $2\sqrt{2}$,2) and (2,2,2) supercells of STO with respect the primitive high symmetry *Pm-3m* cubic cell – 4 and 8 formula units, respectively. Projector Augmented Wave potentials [36] are used with only treating *d* states as valence electrons for Ti cations. Wannier functions are built using the wannier90 package [26,27,29,34,37].

*Strain phase diagram:* **Figure 1.b** displays the total energy difference of the most stable structures relaxed with DFT with respect to the high symmetry *P4/mmm* structure – the high symmetry primitive cell for epitaxially strained perovskites – as a function of the substrate lattice parameter. At low compressive strain ($|\eta|$<1.35%), STO adopts a centrosymmetric *I4/mcm* cell very similar to the bulk material. This phase is characterized by a $a^0a^0c^-$ octahedral rotation pattern. Upon increasing the compressive strain between 1.35%< $|\eta|$<5.58%, a ferroelectric (FE) phase is identified adopting a *I4cm* cell and displaying an additional polar axis along z with respect to the *I4/mcm* cell (see sketches in **Fig1.a**). This result is reminiscent of the experimental observation of ferroelectricity in SrTiO$_3$ under a moderate compressive or tensile strain [18]. Finally, at a compressive strain of 6% ($a_{sub}$=3.675 Å), STO adopts a FE *P4mm* tetragonal structure (**Fig. 1.a**) characterized by a polar axis along z and the total absence of any octahedral rotations. Surprisingly, a strong tetragonality with a *c/a* ratio of 1.29 is observed within this phase, while the structure stabilized at $a_{sub}$=3.69 Å has a tetragonality of 1.19. This results hints at the "T-phase" stabilized for high compressive strain in BiFeO$_3$ [38,39] or in BaTiO$_3$ and PbTiO$_3$ under pressure [40–42]. It also relates with the ferroelectric phases predicted in perovskites undergoing high tensile strains [43,44]. These results suggest that a super-ferroelectric phase could be generic for perovskites experiencing large tensile and/or compressive strains.

**Figure 1.c** reports the $a^0a^0c^-$ octahedral rotation and polar mode amplitudes extracted from a symmetry mode analysis with respect to a highly symmetric undistorted *P4/mmm* cell as well as the computed spontaneous polarization as a function of the strain values. The spontaneous polarization – or polar mode amplitude – monotonously increases from 8.3 µC.cm$^{-2}$ ($|\eta|$=1.35%) to 41.7 µC.cm$^{-2}$ ($|\eta|$=5.58%) upon increasing the compressive strain value once the FE *I4cm* phase is reached. Within the super tetragonal *P4mm* phase, a giant polarization of 97 µC.cm$^{-2}$ is reached, again reminiscent of the situation reached in the T-phase of BiFeO$_3$ (P=150 µC.cm$^{-2}$ [39]). The $a^0a^0c^-$ octahedral rotation amplitude also increases with increasing the strain value, suggesting that ferroelectricity and octahedral rotations are not strongly antagonists in strained SrTiO$_3$ [16,45]. This is consistent with Ref. [46] that previously showed that AFD motions and ferroelectricity can cooperate in SrTiO$_3$ once a sufficiently large tetragonality of the cell is achieved. These results are insensitive to the choice of the DFT exchange-correlation functional as the GGA Perdew Burke Ernzerhof revised for solid (PBEsol) DFT functional yields the very same trend with epitaxial strain (see **supplementary material 2**).

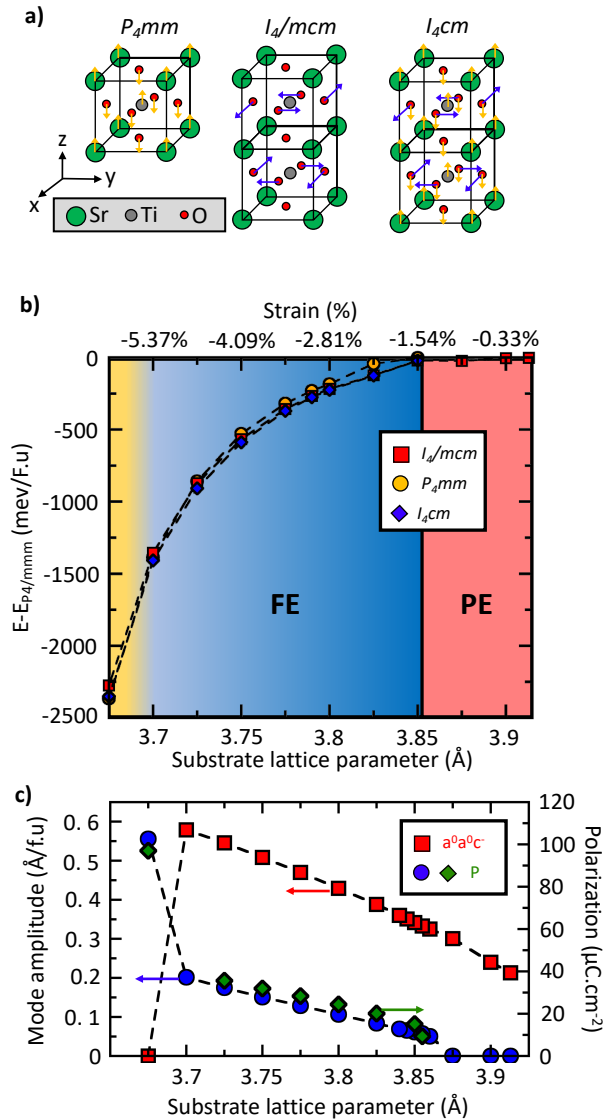

*Figure 1: Phase diagram of SrTiO$_3$ under compressive strain. a) Sketches of the structural motion associated with the polar mode, the $a^0a^0c^-$ antiferrodistortive and the combination of polar and antiferrodistortive motions yielding the P$_4$mm, I$_4$/mcm and I$_4$cm cells, respectively. Arrows are not representative of the actual displacement amplitude appearing in the compound. b) Energy difference (in meV/f.u) of most stable phases with respect to the high symmetry undistorted P$_4$/mmm cell of SrTiO$_3$ as a function of the strain. c) Amplitudes (in Å/f.u) of polar (blue filled circles) and $a^0a^0c^-$ octahedral*

rotation distortions (red filled squares) extracted from a symmetry mode analysis with respect to the high symmetry undistorted $P_4/mmm$ cell (left scale) and computed spontaneous polarization (in µC.cm$^{-2}$, green filled diamonds, right scale) for the identified ground states as a function of the strain.

**Rashba parameters as a function of the polar mode amplitude:** Within the ferroelectric phases, STO lacks an inversion center since ions undergo polar displacements. This is similar to the situation appearing in 2-DEG formed in STO when a local symmetry breaking occurs at its surface and/or interface with other compounds. Hence, it can yield a Rashba spin-orbit effect with a spin-splitting of the Ti- $t_{2g}$ bands located at the bottom of the conduction band (CB) [15]. **Figure 2.a** displays the band structure along the $\Gamma - X$ path associated with the bottom of CB for the $I_4cm$ phase reached at 2.81% of compressive strain, this including the SOI. The degeneracy of bands at the bottom of CB is lifted according to the spin flavor $\pm S_y$, no $\pm S_x$ spin dependance is found for these bands – it is observed for bands dispersing along the $\Gamma - Y$ path (**Fig. 2.c**). Thus, a spin locking orthogonal to the momentum k of electrons and to the polar axis is identified, confirming the existence of Rashba phenomenon in polar STO. Finally, as observed in Ferroelectric Rashba materials [13,14,47–50], a non-volatile control of the spin locking is reached as switching the polarization reverses the band splitting (**Figures 2.b** and **d**).

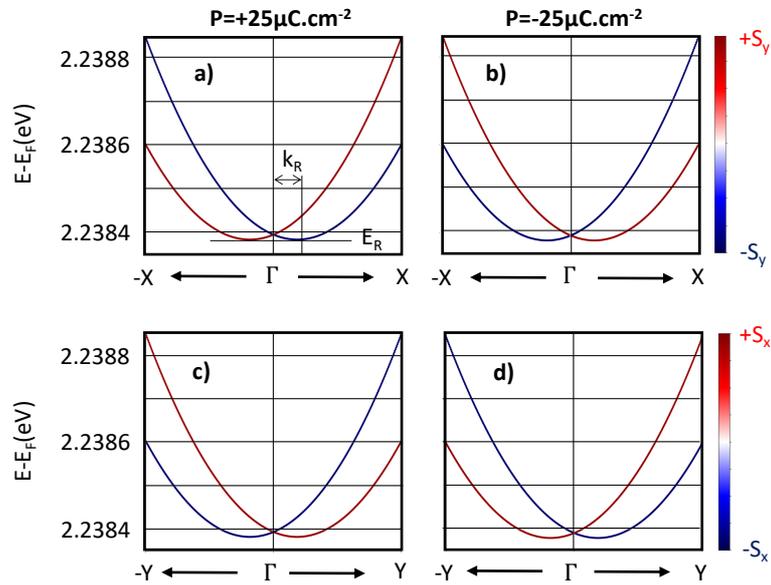

*Figure 2: Rashba spin splitting at the bottom of the conduction band in ferroelectric SrTiO$_3$.* Bands dispersion along the $\Gamma$-X *(a,b)* and $\Gamma$-Y *(c,d)* paths projected on the $S_x$ *(c,d)* and $S_y$ *(a,b)* spin components for states at the bottom of the conduction band for a polarization up *(a,c)* or down *(b,d)*. The ferroelectric state reached with a substrate lattice parameter of 3.80 Å is considered here. High symmetry points are $\Gamma$ (0,0,0), X(½,0,0) and Y (0,½,0) and refers to the $I_4cm$ cell Brillouin zone.

Following the confirmation of the existence of Rashba states at the bottom of the conduction band of FE STO, the Rashba parameter $\alpha_R$ is extracted from DFT band structure calculations using the following formula $\alpha_R = \frac{2E_R}{k_R}$ where $E_R$ is the energy splitting with respect to the high symmetry positions and $k_R$ is the momentum offset (see **Fig. 2.a**). **Figures 3.a** and **b** summarize the evolution of $\alpha_R$ as a function of the polar mode and strain amplitudes, respectively. The Rashba coefficient $\alpha_R$ is estimated between ten to a hundred of meV.Å, an order of magnitude in agreement with the 30 meV.Å estimated experimentally for the 2-DEG appearing at the LaAlO$_3$/SrTiO$_3$ interface. Nevertheless, $\alpha_R$ monotonously decreases with the polar mode amplitude and even reaches 0 within the super-tetragonal phase despite the latter phase shows the largest spontaneous polarization (P=97μC.cm$^{-2}$). By looking at the evolution of $\alpha_R$ with the substrate lattice parameter (**Fig 3.b**), $\alpha_R$ in fact diverges and reaches a maximal value at the boundary between the ferroelectric-paraelectric (FE-PE) phases reached under epitaxial strain. Such a behavior is surprising since the Rashba coefficient should *a priori* scales linearly with the polar displacement amplitude.

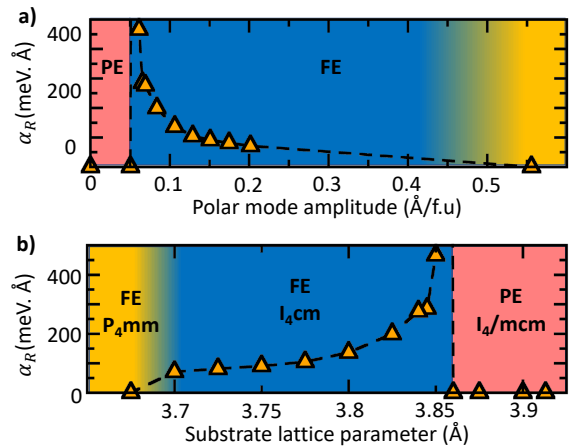

*Figure 3: Rashba parameter as a function of strain and polar displacements amplitude. Computed Rashba parameter $\alpha_R$ (in meV.Å) as a function of polar displacement amplitude (in Å/f.u, panel **(a)** and substrate lattice parameter (in Å, panel **b**) in strained SrTiO$_3$ films.*

***The Rashba parameter is dominated by polar displacements amplitude in two different regimes:*** In order to understand the surprising behavior of $\alpha_R$ as a function of epitaxial strain/polar mode amplitude, it is important to disentangle the contribution of the distortions appearing in the materials on $\alpha_R$, namely octahedral rotations and polar displacements. To that end, the calculation of $\alpha_R$ is performed within a perfectly undistorted cubic cell of SrTiO$_3$ as a function of the polar and AFD mode amplitude Q$_P$ and Q$_{AFD}$, respectively. Results are summarized in **Fig. 4.a**. Starting from the cubic cell with only the polar mode condensed, $\alpha_R$ first exhibits a linear trend as a function of Q$_P$ for Q$_P$<0.025 Å/f.u and then it becomes inversely proportional to Q$_P$ (i.e $\alpha_R \propto 1/Q_P$) for Q$_P$>0.025 Å/f.u. Adding a fixed octahedral rotations within the cubic cell with an amplitude $Q_{AFD}$=0.210 Å/f.u, the trend and values of $\alpha_R$ as a function of $Q_P$ are not substantially modified. The minor role of octahedral rotations is confirmed by performing the calculation of $\alpha_R$ as a function of $Q_{AFD}$ at fixed polar displacement amplitude $Q_P$=0.104 Å/f.u that shows a nearly constant $\alpha_R$ – it only increases by 12% between $Q_{AFD}$=0 Å/f.u and $Q_{AFD}$=0.420 Å/f.u. Therefore, one concludes here that polar displacements are the key factor behind the trend of $\alpha_R$ versus the epitaxial strain.

***The polar displacements alleviate $t_{2g}$ states degeneracies and annihilate the Rashba phenomena:*** The next step is to understand the two $Q_P$ ranges producing either a proportional or inversely proportional trend between $\alpha_R$ and $Q_P$. In a perfectly undistorted cell of SrTiO$_3$, the three $t_{2g}$ states are perfectly degenerate. However, the lattice distortions can introduce a crystal field splitting $\Delta_{CF}$ alleviating the degeneracy of the $t_{2g}$ states. **Figure 4.b** displays the evolution of $\Delta_{CF}$= $E_{dxz/dyz}$-$E_{dxy}$ as a function of the different lattice mode amplitude where $E_{dxy, dxz, dyz}$ are the on-site energies of the $t_{2g}$ orbitals obtained by a construction of atomic-like Wannier Functions centered on Ti cations – only the Kohn-Sham states associated with $t_{2g}$ states are considered for the construction, *i.e.* 12 states for a 4 f.u of STO. Firstly, polar displacements produce a strong $\Delta_{CF}$ with $\Delta_{CF} = 6314 Q_P^2$, whose evolution is not altered by the introduction of AFD motions – only the crystal field at zero mode amplitude $\Delta_{CF0}$ is shifted to higher values by the introduction of AFD motion – while octahedral rotation brings a contribution to $\Delta_{CF}$ that is one order of magnitude smaller with $\Delta_{CF}$-$\Delta_{CF0}$= $395 Q_{AFD}^2$. Secondly, similarly to $\alpha_R$, one recovers two distinct regimes for $\Delta_{CF}$ as a function of $Q_P$: (i) $\Delta_{CF}$ is constant for $Q_P$<0.025 Å/f.u (ii) while it becomes sizable and large for $Q_P$>0.025 Å/f.u. Possessing two regimes as a function of $Q_P$, $\Delta_{CF}$ and $\alpha_R$ may be connected together.

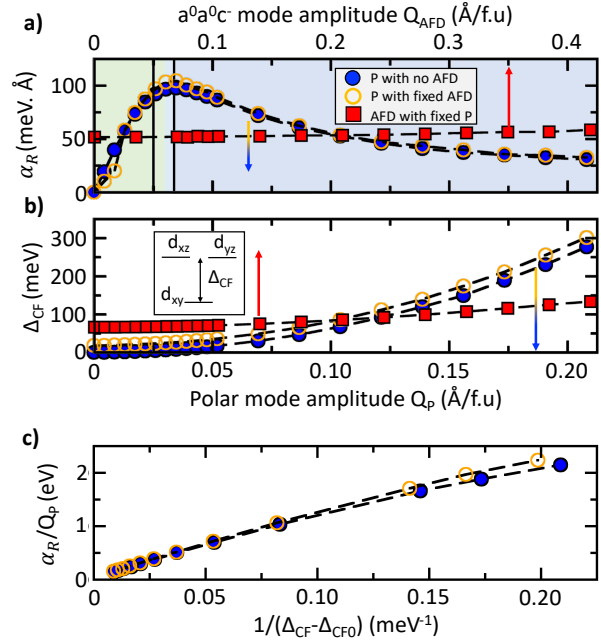

*Figure4: Rashba parameter as a function of lattice distortions in a cubic SrTiO$_3$. a-b) Computed Rashba parameter (**a**, in meV.Å, upper panel) and crystal field splitting Δ$_{CF}$ (**b**, meV, lower panel) between the d$_{xy}$ and d$_{xz}$/d$_{yz}$ orbitals as a function of the polar mode amplitude Q$_P$ (in Å/f.u, lower scale) in a cubic cell of SrTiO$_3$ with (orange unfilled circles) and without (blue filled circles) a$^0$a$^0$c$^-$ AFD motions and as a function of AFD motion amplitude Q$_{AFD}$ (in Å/f.u, upper scale, red filled squares) at a fixed polar amplitude. The amplitude of the frozen AFD motion is 0.296 Å/f.u (orange unfilled circles) and of the frozen polar mode amplitude is 0.104 Å/f.u (red filled squares). Only the states at the bottom of the conduction band are considered for extracting the Rashba parameter. **c)** Normalized Rashba parameter $\alpha_R/Q_P$ (meV) as a function of the inverse of the crystal field splitting induced by the polar displacement. Δ$_{CF0}$ is the crystal field splitting appearing at Q$_P$=0 Å.*

**$\alpha_R$ is inversely proportional to Δ$_{CF}$:** **Figure 4.c** displays the evolution of $\alpha_R/Q_P$ as a function of 1/(Δ$_{CF}$-Δ$_{CF0}$) for the Q$_P$>0.0025 Å/f.u regime. A clear linear trend between the inverse crystal field splitting induced by polar displacements and the normalized Rashba parameter is identified, either in the presence or not of octahedral rotation. Thus, although $\alpha_R$ is proportional to Q$_P$ by construction of the SOI, it is inversely proportional to the crystal field splitting of states undergoing the Rashba phenomena. The latter observation is compatible with the results of Bahramy *et al* who showed that the amplitude of $\alpha_R$ scales as 1/(E$_m$-E$_n$) where E$_m$-E$_n$ are the energies of states *m* and *n* experiencing the Rashba effects [10]. Nevertheless, the observed effect in SrTiO$_3$ is subtle: (i) in all situations, $\alpha_R$ scales as $Q_P/\Delta_{CF}$ but (ii) at small polar displacement (*i.e.* Q$_P$<0.025 Å/f.u), $\alpha_R$ scales linearly with Q$_P$ since Δ$_{CF}$ is almost constant and (iii) for Q$_P$ >0.025 Å/f.u, the $1/\Delta_{CF} \propto 1/Q_P^2$ contribution to $\alpha_R$ dominates and hence $\alpha_R$ ultimately evolves as $1/Q_P$. This is compatible with the trend of $\alpha_R$ as a function of Q_P of **Fig.4.a.** It further explains the surprising trend of $\alpha_R$ with epitaxial strain of **Fig3.b.** At low compressive strain, the polar displacements are small and $\Delta_{CF}$ is close to zero, hence producing large $\alpha_R$ and even diverging at the paraelectric-ferroelectric boundary. At larger compressive strain, including the super tetragonal phase, polar displacements alleviate $t_{2g}$ states degeneracies and hence annihilate the Rashba-SOI. Finally, it is worth emphasizing that the computed $\alpha_R$ in compressively strained compounds is larger than the one computed in a perfect cubic cell at fixed polar mode amplitude. This is due to an additional crystal field splitting of Ti $t_{2g}$ orbitals induced by the compressive strain that promotes the stabilization of the $d_{xz}/d_{yz}$ orbitals over the $d_{xy}$ orbital, in opposition to the polar mode, hence diminishing the overall crystal field acting on Ti $t_{2g}$ states.

**Conclusion:** Although $\alpha_R$ is often expected to monotonously scale with the polar mode amplitude by construction of the SOI, the polar displacements can alleviate the degeneracy of the Ti $t_{2g}$ states and quench the Rashba-SOI. Consequently, the Rashba parameter is bound to a upper value. Preserving the Ti $t_{2g}$ orbital degeneracies over searching for larger electrical dipoles is thus the key to achieve large Rashba parameters in STO. Although applied in this Letter to ferroelectric phases of STO reached under compressive epitaxial strain, the concept is in fact general to any material showing polar displacements of atoms undergoing the Rashba-SOI. For instance, one may consider playing with epitaxial strain in order to modify the Ti $t_{2g}$ states degeneracies appearing in the 2-DEG for reaching more efficient spin-to-charge current interconversions [15].

**Acknowledgements**

JV acknowledge discussion with M. Bibes, L. Vila and J.P. Attané. JV acknowledges access granted to HPC resources of Criann through the projects 2020005 and 2007013 and of Cines through the DARI project A0080911453. This work was supported by the French ANR through the project "CONTRABASS".